\documentclass[preprint,aps,floats,prb,fancyhdr,a4paper,superscriptaddress,nofootinbib,showpacs]{revtex4}
\usepackage{latexsym}
\usepackage{amsmath}
\usepackage{amssymb}
\usepackage{amsfonts}
\usepackage{graphicx}
\usepackage{dcolumn}
\usepackage{fancyhdr}
\usepackage[colorlinks]{hyperref}

\begin{document}

\title{FULL CAUSAL BULK VISCOUS COSMOLOGIES WITH TIME-VARYING CONSTANTS}

\author{\footnotesize JOS\'E ANTONIO BELINCH\'{O}N\footnote{
e-mail: abelcal@ciccp.es}}

\address{Grupo Inter-Universitario de An\'{a}lisis Dimensional. \\
Dept. F\'{\i}sica ETS Arquitectura UPM Av. Juan de Herrera 4. Madrid
28040 Espa\~{n}a.}

\author{\footnotesize INDRAJIT CHAKRABARTY}

\address{Blk 278, \#06-183, Toh Guan Road, Singapore 600278}

\begin{abstract}
We study the evolution of a flat Friedmann-Robertson-Walker Universe, filled
with a bulk viscous cosmological fluid, in the presence of time varying
``constants''. The dimensional analysis of the model suggests a
proportionality between the bulk viscous pressure of the dissipative fluid
and the energy density. On using this assumption and with the choice of the
standard equations of state for the bulk viscosity coefficient, temperature
and relaxation time, the general solution of the field equations can be
obtained, with all physical parameters having a power-law time dependence.
The symmetry analysis of this model, performed by using Lie group
techniques, confirms the unicity of the solution for this functional form of
the bulk viscous pressure. In order to find another possible solution we
relax the hypotheses assuming a concrete functional dependence for the
``constants''.
\end{abstract}
\maketitle
\section{Introduction}

Since the pioneering work of Dirac$^{1}$, who proposed, motivated by
the occurrence of large numbers in Universe, a theory with a time variable
gravitational coupling constant $G$, cosmological models with variable $G$
and nonvanishing and variable cosmological term have been intensively
investigated in the physical literature$^{2-14}$.

Dissipative thermodynamic processes in cosmology originating from a bulk
viscosity are believed to play an important role in the dynamics and
evolution of the Universe. Misner$^{15}$ suggested that large-scale
isotropy of the universe observed at the present epoch is due to the action
of neutrino viscosity which was not negligible when the universe was less
than a second old. There are a number of processes responsible for producing
bulk viscosity in the early universe, such as the interaction between
radiation and matter$^{16}$, gravitational string production$^{17-18}$,
viscosity due to quark and gluon plasma, dark matter or particle
creation $^{24, 33-34}$.

In order to study these phenomena, the theories of dissipation in
Eckart-Landau formulation$^{19-20}$, who made the first attempt at
creating a relativistic theory of viscosity. However, these
theories are now known to be pathological in several ways. Regardless of the
choice of equation of state, all equilibrium states in these theories are
unstable. In addition, as shown by Israel$^{21}$, signals may be
propagated through the fluid at velocities exceeding the speed of light in
contradiction with the principle of causality. These problems arise due to
the first order nature of the theory since it considers only first-order
deviations from the equilibrium, leading to parabolic differential
equations, and hence to infinite speeds of propagation for heat flow and
viscosity. While such paradoxes appear particularly glaring in relativistic
theory, infinite propagation speeds already constitutes a difficulty at the
classical level, since one does not expect thermal disturbances to be
carried faster than some (suitably defined) mean molecular speed.
Conventional theory is thus applicable only to phenomena which are
``quasi-stationary'' i.e. slowly varying on space and time scales
characterized by mean free path and mean collision time$^{21}$. This is
inadequate for many phenomena in high-energy astrophysics and relativistic
cosmology involving steep gradients or rapid variations. These deficiencies
can be traced to the fact that the conventional theories (both classical and
relativistic) make overly restrictive hypothesis concerning the relation
between the fluxes and densities of entropy, energy and particle number.

A relativistic second-order theory was found by Israel$^{21}$ and
developed by Israel and Stewart$^{22}$ into what is called
`transient' or \ `extended' irreversible thermodynamics. In this model,
deviations from equilibrium (bulk stress, heat flow and shear stress) are
treated as independent dynamical variables leading to a total of $14$
dynamical fluid variables to be determined. The solutions of the full causal
theory are well behaved for all times. Hence the advantages of the causal
theories are the following$^{23}$: 1) for stable fluid
configurations, the dissipative signals propagate causally 2) unlike
Eckart-type's theories, there is no generic short-wavelength secular
instability in causal theories and 3) even for rotating fluids, the
perturbations have a well-posed initial value problem. Therefore, the best
currently available theory for analyzing dissipative processes in the
Universe is the full Israel-Stewart causal thermodynamics.

Due to the complicated nonlinear character of the evolution equations, very
few exact cosmological solutions of the gravitational field equations are
known in the framework of the full causal theory. For a homogeneous Universe
filled with a full causal viscous fluid source obeying the relation $\xi
\sim\rho^{\frac{1}{2}}$, exact general solutions of the field equations have
been obtained earlier$^{24-33}$. In this case the evolution of the bulk
viscous cosmological model can be reduced to a Painleve-Ince type
differential equation. It has also been proposed that causal bulk viscous
thermodynamics can model on a phenomenological level matter creation in the
early Universe$^{24-33}$.

In this paper, we consider the evolution of a causal bulk viscous fluid
filled flat FRW type Universe, by assuming the standard equations of state
for the bulk viscosity coefficient, temperature and relaxation time and in
the presence of time varying constants ($G, c$ and $\Lambda)$. In order to
obtain some very general properties of this cosmological model with variable
constants, we shall adopt a method based on the studies of the symmetries of
the field equations. As a first step, we shall study the field equations
from dimensional point of view. The dimensional method provides general
relations between physical quantities and allows us to make some definite
assumptions on the behavior of thermodynamical quantities as well as on the
equation of state for the bulk viscous parameter. In particular, we find
that, under the assumption of the conservation of the total energy of the
Universe, the bulk viscous pressure of the cosmological fluid must be
proportional to the energy density of the matter component and for $%
\gamma=1/2$ (where $\gamma$ stands for the bulk viscous parameter), we find
that $G/c^{2}$ must remain constant in spite of considering both
``constants'' as functions of time $t.$ On using this assumption, the
gravitational field equations can be integrated exactly, leading to a
general solution in which all thermodynamical quantities have a power-law
time dependence.

As we have been able to find a solution through dimensional analysis, it is
possible that there are other symmetries of the model, since dimensional
analysis is a reminiscent of scaling symmetries, which obviously are not the
most general form of symmetries. Hence, we shall study the model through the
method of Lie group symmetries, showing \ that under the assumed hypotheses
of the proportionality of bulk viscous pressure to the energy density, there
are no other solutions of the field equations.

The paper is organized as follows: In section (\ref{M}), we outline the
equations of the model, define all the quantities as well as their equations
of state and fix the notation. In section (\ref{H}), we study our equations
from the dimensional point of view (subsection \ref{dym}). This study allows
us to make two simplifying hypotheses which we use to integrate the
equations. In subsection (\ref{STA}), we obtain the equation of state for
the bulk viscosity through the dynamical approach. Section (\ref{DA})
presents a naive method to study our model taking into account the previous
hypotheses. This technique allows us to obtain a complete set of solutions
as well as to arrive at some interesting conclusions. In section (\ref{INT}%
), we check the solutions obtained in the previous sections are correct and
show that direct integration does not give more information about the
solutions. As Dimensional Analysis is a manifestation of symmetries, in
section (\ref{LIE}) we show that under the imposed hypothesis there are no
more solutions for the field equations than ones already obtained. In
section (\ref{REL}), we try to relax the hypotheses in order to seek other
solutions but (un)fortunately we only recover our previous solutions. We end
the paper by summarising our findings in section (\ref{CON}).

\section{The Model\label{M}}

Following Maartens$^{35}$, we consider a Friedmann-Robertson-Walker (FRW)
Universe with a line element
\begin{equation}
ds^{2}=c^{2}dt^{2}-f^{2}(t)\left( dx^{2}+dy^{2}+dz^{2}\right) ,  \label{line}
\end{equation}
filled with a bulk viscous cosmological fluid with the following
energy-momentum tensor:
\begin{equation}
T_{i}^{k}=\left( \rho+p+\Pi\right) u_{i}u^{k}-\left( p+\Pi\right)
\delta_{i}^{k},  \label{1}
\end{equation}
where $\rho$ is the energy density, $p$ the thermodynamic pressure, $\Pi$ is
the bulk viscous pressure and $u_{i}$ is the four velocity satisfying the
condition $u_{i}u^{i}=1$. The number 4-flux and the entropy 4-flux take the
form:
\begin{eqnarray}
N^{i} & = & nu^{i},  \label{num} \\
S^{i} & = & sN^{i}-\left( \frac{\tau\Pi^{2}}{2\xi T}\right) u^{i},
\label{entropy}
\end{eqnarray}
where $n$ is the number density, $s$ the specific entropy, $T\geq0$ the
temperature, $\xi$ the bulk viscosity coefficient and $\tau\geq0$ the
relaxation coefficient for transient bulk viscous effect (i.e. the
relaxation time). The fundamental thermodynamic tensors
(\ref{1}-\ref{entropy}) are subject to the dynamical laws of
energy-momentum conservation, number conservation and the Gibb's equation:
\begin{eqnarray}
T_{i;k}^{k} & = & 0,  \label{con1} \\
N_{;i}^{i} & = & 0,  \label{con2} \\
Tds & = & d\left( \frac{\rho}{n}\right) +pd\left( \frac{1}{n}\right) .
\label{con3}
\end{eqnarray}

The equations (\ref{1}-\ref{entropy}) and (\ref{con1}-\ref{con3}) imply:
\begin{equation}
TS_{;i}^{i}=-\Pi\left( 3H+\frac{\dot{\tau}}{\xi}\dot{\Pi}+\frac{1}{2}%
T\Pi\left( \frac{\tau}{\xi T}u^{i}\right) _{;i}\right) ,  \label{preequation}
\end{equation}
by equations (\ref{con1}-\ref{con3}) and (\ref{preequation}), the simplest
way (linear in $\Pi)$ to satisfy the $H$-theorem (i.e. for the entropy
production to be non-negative, $S_{;i}^{i}=\frac{\Pi^{2}}{\xi T}\geq0$
$^{22, 36-37}$) leads to the causal evolution equation for bulk
viscosity given by$^{35}$
\begin{equation}
\tau\dot{\Pi}+\Pi=-3\xi H-\frac{\epsilon}{2}\tau\Pi\left( 3H+\frac{\dot{\tau
}}{\tau}-\frac{\dot{\xi}}{\xi}-\frac{\dot{T}}{T}\right) ,  \label{bulk}
\end{equation}

In eq.(\ref{bulk}), $\epsilon=0$ gives \ the truncated theory (the truncated
theory implies a drastic condition on the temperature), while $\epsilon=1$
gives the full theory. The non-causal theory has $\tau=0$.

The growth of the total commoving entropy $\Sigma$ over a proper time
interval $\left( t_{0},t\right) $ is given by$^{35}$:
\begin{equation}
\Sigma(t)-\Sigma\left( t_{0}\right) =-\frac{3}{k_{B}}\int_{t_{0}}^{t}\frac{%
\Pi Hf^{3}}{T}dt,  \label{M entropy}
\end{equation}
where $k_{B}$ is the Boltzmann's constant. The Einstein gravitational
field equations with variable $G$, $c$ and $\Lambda$ are:
\begin{equation}
R_{ik}-\frac{1}{2}g_{ik}R=\frac{8\pi G(t)}{c^{4}\left( t\right) }%
T_{ik}+\Lambda(t)g_{ik}.  \label{ECU1}
\end{equation}
Applying the covariance divergence to the second member of equation (\ref
{ECU1}) we get:
\begin{equation}
div\left( \frac{G}{c^{4}}T_{i}^{j}+\delta_{i}^{j}\Lambda\right) =0,
\label{conser1}
\end{equation}
\begin{equation}
T_{i;j}^{j}=\left( \frac{4c_{,j}}{c}-\frac{G_{,j}}{G}\right) T_{i}^{j}-\frac{%
c^{4}\delta_{i}^{j}\Lambda_{,j}}{8\pi G},  \label{conser2}
\end{equation}
that simplifies to:
\begin{equation}
\dot{\rho}+3\left( \rho+p\right) H+3H\Pi=-\frac{\dot{\Lambda }%
c^{4}}{8\pi G}-\rho\frac{\dot{G}}{G}-4\rho\frac{\dot{c}}{c},  \label{conser3}
\end{equation}
where $H$ stands for the Hubble parameter ({\it{$H=\dot{f}/f$}}$).$ The
last equation may be written in the following form:
\begin{equation}
\dot{\rho}+3\left( \rho+p\right) H+3H\Pi+\frac{\dot{\Lambda}c^{4}%
}{8\pi G}+\rho\frac{\dot{G}}{G}-4\rho\frac{\dot{c}}{c}=0.  \label{def0}
\end{equation}
Therefore, our model (with FRW symmetries) is described by the following
equations:
\begin{eqnarray}
2\dot{H}+ 3 H^{2} & = & -\frac{8\pi G}{c^{2}}\left( p+\Pi\right) +\Lambda c^{2},
\label{field1} \\
3H^{2} & = & \frac{8\pi G}{c^{2}}\rho+\Lambda c^{2},  \label{field2} \\
\dot{\rho}+3\left( \rho+p+\Pi\right) H & = & -\frac{\dot{\Lambda}c^{4}}{8\pi G}%
-\rho\frac{\dot{G}}{G}+4\rho\frac{\dot{c}}{c},  \label{field3} \\
\tau\dot{\Pi}+\Pi & = & -3\xi H-\frac{\epsilon}{2}\tau\Pi\left( 3H+\frac {\dot{%
\tau}}{\tau}-\frac{\dot{\xi}}{\xi}-\frac{\dot{T}}{T}\right) .  \label{field4}
\end{eqnarray}

In order to close the system of equations (\ref{field1}-\ref{field4}) we
have to give the equation of state for $p$ and specify $T$, $\xi$ and $\tau$%
. As usual, we assume the following phenomenological (ad hoc) laws$^{35}$:
\begin{eqnarray}
p & = & \omega\rho,  \label{csi1} \\
\xi & = & k_{\gamma}\rho^{\gamma},  \label{csi2} \\
T & = & D_{\delta}\rho^{\delta},  \label{csi3} \\
\tau & = & \xi\rho^{-1}=k_{\gamma}\rho^{\gamma-1},  \label{csi4}
\end{eqnarray}
where $0\leq\omega\leq1$, and $k_{\gamma}\geq0$, $D_{\delta}\geq0$ are
dimensional constants, $\gamma\geq0$ and $\delta\geq0$ are numerical
constants. Eqs. (\ref{csi1}) are standard in cosmological models whereas the
equation for $\tau$ is a simple procedure to ensure that the speed of
viscous pulses does not exceed the speed of light. These are without
sufficient thermodynamical motivation, but, in absence of better
alternatives, we use these equations and expect that they will at least
provide an indication of the range of possibilities. For the temperature
law, we take, $T=D_{\delta }\rho^{\delta}$, which is the simplest law
guaranteeing positive heat capacity.

In the context of irreversible thermodynamics $p$, $\rho$, $T$ and the
particle number density $n$ are equilibrium magnitudes which are related by
equations of state of the form $\rho=\rho(T,n)$ and $p=p(T,n)$. From the
requirement that the entropy is a state function, we obtain the equation

\begin{equation}
\left( \frac{\partial\rho}{\partial n}\right) _{T}=\frac{p+\rho}{n}-\frac
{T}{n}\left( \frac{\partial p}{\partial T}\right) _{n}.  \label{thermo1}
\end{equation}
For the equations of state (\ref{csi1}-\ref{csi4}) this relation imposes the
constraint $\delta=\frac{\omega}{\omega+1}$ so that $0\leq\delta\leq1/2 $
for $0\leq\omega\leq1$, a range of values which is usually considered in the
physical literature$^{38}$.

The Israel-Stewart-Hiscock theory is derived under the assumption that the
thermodynamical state of the fluid is close to equilibrium, that is, the
non-equilibrium bulk viscous pressure should be small when compared to the
local equilibrium pressure $\left| \Pi\right| <<p=\omega\rho$ $^{39-40}$.
If this condition is violated then one is effectively
assuming that the linear theory holds also in the nonlinear regime far from
equilibrium. For a fluid description of the matter, this condition should be
satisfied.

Therefore, with all these assumptions and taking into account the
conservation principle, i.e., $div(T_{i}^{j})=0$, the resulting field
equations are as follows:

\begin{eqnarray}
2\dot{H}+ 3 H^{2} & = & -\frac{8\pi G}{c^{2}}\left( p+\Pi\right) +\Lambda c^{2},
\label{nfield1} \\
3H^{2} & = & \frac{8\pi G}{c^{2}}\rho+\Lambda c^{2},  \label{nfield2} \\
\dot{\rho}+3\left( \omega+1\right) \rho H & = & -3H\Pi,  \label{nfield3} \\
\frac{\dot{\Lambda}c^{4}}{8\pi G}+\rho\frac{\dot{G}}{G}-4\rho\frac{\dot{c}}{c
} & = & 0,  \label{nfield4} \\
\dot{\Pi}+\frac{\Pi}{k_{\gamma}\rho^{\gamma-1}} & = & -3\rho H-\frac{1}{2}
\Pi\left( 3H-W\frac{\dot{\rho}}{\rho}\right) ,  \label{nfield5}
\end{eqnarray}
where $W=\left( \frac{2\omega+1}{\omega+1}\right) $.

\section{Hypotheses and equations of state\label{H}}

In this section, we study the field equations in a dimensional way as well
as with a dynamical systems approach in order to obtain relationships
between various physical quantities, simplify the field equations and to
obtain an adequate equation of state for the bulk viscous parameter.

We use dimensional considerations to re-write the field equations in a
dimensionless way and show that some interesting relations can be obtained
as a result. We find that the bulk viscous pressure has the same behaviour
as the energy density and if $\gamma\neq1/2$, the relationship $G/c^{2}$
should vary if we demand that our equations remain gauge and scale
invariant. We find that with the special case $\gamma=1/2$ the relationship $%
G/c^{2}=const. $ but in such a way that both ``constants'' vary. Dimensional
analysis suggests that $\gamma=1/2$ is a very special value for the bulk
viscous parameter. Using the dynamical systems approach, we shall show that
for this value of $\gamma$, our model approximates the dynamics of a perfect
fluid and for large time, the dynamics of the Universe follows that of a
flat FRW model.

\subsection{Dimensional considerations\label{dym}}

The $\pi-monomia$ is the main object in dimensional analysis. It may be
defined as the product of quantities which are invariant under change of
fundamental units. $\pi-monomia$ are dimensionless quantities, their
dimensions are equal to unity. The dimensional analysis has structure of Lie
group$^{41-44}$. The $\pi-monomia$ are invariants under the action of
the similarity group. We must mention here that the similarity
group is only a special class of the group of all symmetries that can be
obtained using the Lie method. For this reason, when one uses dimensional
analysis only one of the several possible solutions to the problem is
obtained.

The equations (\ref{nfield1}-\ref{nfield5}) and the equation of state (\ref
{csi1}-\ref{csi4}) can be expressed in a dimensionless way by the following $%
\pi-monomia$:
\begin{eqnarray}
\pi_{1} & = & \frac{Gpt^{2}}{c^{2}} \mbox{\ \ \ \ \ \ } \pi_{2} = \frac{G\Pi
t^{2}}{c^{2}} \mbox{\ \ \ \ \ \ } \pi_{3} = \frac{G\rho t^{2}}{c^{2}}
\label{pi1} \\
\pi_{4} & = & \frac{\Pi}{p}\mbox{\ \ \ \ \ \ \ \ \ \ } \pi_{5} = \frac{\xi }{\Pi t}
\mbox{\ \ \ \ \ \ \ \ }\pi_{6} = \frac{\tau}{t} = \tau H  \label{pi2} \\
\pi_{7} & = & \frac{\xi}{k_{\gamma}\rho^{\gamma}}\mbox{\ \ \ \ \ \ }
\pi_{8} = \frac{\xi}{\tau\rho}\mbox{\ \ \ \ \ \ \ \ }\pi_{9} = \frac{T}{D_{\delta}\rho^{\delta}}
\label{pi3} \\
\pi_{10} & = & \frac{\rho}{p}\mbox{\ \ \ \ \ \ \ \ \ }\pi_{11} = \Lambda c^{2}t^{2}
\mbox{\ \ \ \ \ \ }\pi_{12} = \frac{\Lambda c^{4}}{G\rho}  \label{pi4}
\end{eqnarray}

The following relations can be obtained from the $\pi-monomia.$

\begin{enumerate}
\item

From $\pi_{1},\pi_{2},\pi_{3},\pi_{4}$ and $\pi_{10}$, we see that
\begin{equation}
\rho\propto p\propto\Pi,
\end{equation}
and thus, $\left[ \Pi\right] =\left[ p\right] =\left[ \rho\right] $.
Therefore, these physical quantities have the same dimensional equation and
exhibit the same behaviour in order of magnitude.

\item

On using $\pi_{7}$ and $\pi_{8}$, we obtain the following relation

\begin{equation}
\widetilde{\pi}_{8}=\frac{k_{\gamma}\rho^{\gamma-1}}{t}
\Longrightarrow\rho=\left( k_{\gamma}^{-1}t\right) ^{b}  \label{edensity}
\end{equation}
and from $\pi_{3}$ and $\widetilde{\pi}_{8}$%
\begin{equation}
\frac{G\rho t^{2}}{c^{2}}=\frac{k_{\gamma}\rho^{\gamma-1}}{t}
\Longrightarrow\frac{c^{2}}{Gt^{2}}=\left( \frac{t}{k_{\gamma}}\right)
^{1/\left( \gamma-1\right) }\Longrightarrow \frac{%
k_{\gamma}^{b}c^{2}}{G}=t^{b+2}  \label{cG}
\end{equation}
where $b=1/(\gamma-1)$. When $\gamma=1/2$, we obtain the relationship $%
k_{\gamma}^{2}=c^{2}/G,$ which indicates that these relations remain
constant, i.e., $G$ and $c$ can vary but in such a way that $c^{2}/G$
remains constant$.$ If $\gamma\neq1/2$, the ``constants'' $G$ and $c$ must
vary if we want that our equations to remain gauge invariant or scale
invariant. In a reasonable physical approach, we need to impose the
condition $G/c^{2}=constant$, and in the process, uniquely fixing the value
of the coefficient $\gamma$. The dimensional analysis not only fixes the
value of $\gamma$, but also predicts the value of the proportionality
coefficient as being $k_{\gamma}^{2}=c^{2}/G$ (in natural units $%
k_{\gamma}^{2}=8\pi$).

\end{enumerate}

Therefore, the solutions that Dimensional Analysis suggests us are the
following:
\begin{eqnarray}
\rho & \propto & \left( k_{\gamma}^{-1}t\right) ^{b}, \\
p & = & \omega\rho\propto\left( k_{\gamma}^{-1}t\right) ^{b}, \\
\Pi & \propto & \left( k_{\gamma}^{-1}t\right) ^{b}=\kappa\rho, \\
\Lambda & \propto & c^{-2}t^{-2}, \\
\frac{G}{c^{2}} & \propto & k_{\gamma}^{b}t^{-b-2},
\end{eqnarray}
where $\kappa$ is a numerical constant. From the equation of state, we
obtain
\begin{eqnarray}
\xi & = & k_{\gamma}\rho^{\gamma}\propto k_{\gamma}\left(
k_{\gamma}^{-1}t\right) ^{b\gamma}, \\
T & = & D_{\delta}\rho^{\delta}\propto\left( k_{\gamma}^{-1}t\right) ^{\delta b}%
\mbox{   with    }  \delta=\frac{\omega}{\omega+1}, \\
\tau & = & \xi\rho^{-1}=k_{\gamma}\left( k_{\gamma}^{-1}t\right) ^{b\left(
\gamma-1\right) }, \mbox{  i. e.  } \tau=t
\end{eqnarray}

So far, we have seen that with this simple method we can obtain the
behaviour of all the physical quantities. We would like to emphasize that we
only have compared $\pi-monomia$ and used the equation of state to obtain
these results. In section \ref{DA}, we will present a naive way to work with
another dimensional method.

\subsection{Structural stability of bulk viscous cosmological models\label{STA}}

In this subsection, we consider the analysis of bulk viscous cosmological
models using dynamical systems theory. Dynamical systems have already been
used in the study of causal viscous fluids$^{28-30}$,
non-causal viscous fluids $^{45}$, $G_{2}$ cosmologies $^{46-47}$
and magnetic fields in scalar field cosmology $^{48}$ (for a
review of dynamical systems in cosmology, see $^{49-50}$). In this method,
the governing equations of the model are
considered as a finite system of autonomous ordinary differential equations.
By using this method, we can determine the arbitrary exponent in the state
equation of the bulk viscosity coefficient.

As a basic physical requirement, we assume that the causal bulk viscous
model approximates the dynamics of a perfect fluid and for large time, the
dynamics of the Universe follows that of a flat FRW model. Our assumption is
supported by the fact that the viscous pressure will decay faster than the
thermodynamic pressure $^{39}$ and hence the behaviour of our model
is equivalent to that of a model with a perfect fluid. On a large scale, the
Universe is well described by the classical flat FRW models. It is natural
to assume here that perturbations introduced by a dissipative parameter like
bulk viscosity does not modify the cosmological evolution even at early
times. Viscous dissipative terms can significantly modify the
thermodynamical behaviour of the early cosmological fluid, but, we do not
expect a major modification in the scale factor. With these assumptions, we
can use mathematical techniques of dynamical systems theory to select a
correct equation of state for the bulk viscosity coefficient, i.e. the
values of the parameter $\gamma$, in such a way that the viscous model
describes a homeomorphic dynamics of the flat FRW model. A natural
environment to study the first order autonomous ordinary differential
equations
\begin{equation}
\frac{dx}{dt}=P(x,y,z),\frac{dy}{dt}=Q(x,y,z),\frac{dz}{dt}=R(x,y,z),
\label{dyna1}
\end{equation}
is a three-dimensional differential manifold $M$ to every point $p=(x,y,z)$
of which the tangent space $T_{p}M$ is associated. The vector field $X\in%
\mathfrak{X}(M)$ $/$ $X(p)=(P(p),Q(p),R(p))\in T_{p}M$ of class $\mathcal{C}%
^{1}$. The set $\mathfrak{X}(M)$ of all such vector fields on $M$ is called
the space of differential equations on $M$. The solution curves, $%
\phi_{t}(p) $ (also called \emph{phase trajectories}) of a vector field $%
X(p) $ on $M$ define a one-parameter group of transformations $\phi
_{t}:M\rightarrow M$ where $t\in\left( a,b\right) .$ If $M$ is compact, then
$t\in\mathbb{R}$ and $\phi_{t}$ is called the \emph{dynamical system}.

Two vector fields $X,Y\in\mathfrak{X}(M)$ are said to be topologically
equivalent if there is a homeomorphism $h:M\rightarrow M$ preserving the
orientation of the solution curves of $X$ into those of $Y$. Let $D$ $%
\subset M$ be a compact subset of $M$. We say that $X\in\mathfrak{X}(D) $ is
\emph{structurally stable} if there is a neighborhood $N(X)$ of $X$ in $%
\mathfrak{X}(D)$ such that for every $Y\subset N(X)$, $Y$ is topologically
equivalent to $X$, i.e. there is an orientation preserving homeomorphism
transforming the phase trajectories of $X$ into those of $Y$.

In the following, we seek spatially homogeneous and isotropic bulk viscous
solutions of the gravitational field equations which, in a structurally
stable way, approximate the dynamics of ordinary (perfect fluid) FRW models.
More precisely, we will look for such spatially and isotropic solutions
which, after being perturbed by the dissipative parameter, yield the
dynamics topologically equivalent to those of FRW models. If we take into
account the state equations (\ref{csi1}-\ref{csi4}), the gravitational field
and the bulk viscous pressure evolution equations can be rewritten in the
following form:
\begin{eqnarray}
\dot{H} & = & -H^{2}-\frac{\left( 3\left( \omega+1\right) -2\right) }{6}\rho-%
\frac{1}{2}\Pi+\frac{1}{3}\Lambda,  \label{dym1} \\
\dot{\rho} & = & -3H\left( \left( \omega+1\right) \rho+\Pi\right) ,
\label{dym2} \\
\dot{\Pi} & = & -\left( \frac{\Pi}{k_{\gamma}\rho^{\gamma-1}}+3\rho H+\frac
{1}{2}\Pi\left[ 3H-W\frac{\dot{\rho}}{\rho}\right] \right) ,  \label{dym3}
\end{eqnarray}
with
\begin{equation}
H^{2}=\frac{1}{3}\rho-\frac{^{3}R}{6}+\frac{1}{3}\Lambda.  \label{dym4}
\end{equation}
Solving
\begin{equation}
\dot{H}=0,\mbox{ \ \ }\dot{\rho}=0,\mbox{ \ \ \ \ }\dot{\Pi}=0,
\label{fixpoint1}
\end{equation}
for $H,\rho$ and $\Pi$, we obtain from (\ref{dym2})
\begin{eqnarray}
\Pi=-(\omega+1)\rho, \nonumber
\end{eqnarray}
and taking into account this result, on using (\ref{dym1}) we arrive at:
\begin{equation}
H^{2}=\frac{1}{3}\left( \rho+\Lambda\right) .  \label{fixpoint2}
\end{equation}

With the use of Eq. (\ref{fixpoint2}), the energy density $\rho$ can be
obtained from the equation (note that $\dot{\rho}=0$)

\begin{equation}
\dot{\Pi}=-\left( \frac{\Pi}{k_{\gamma}\rho^{\gamma-1}}+3\rho H+\frac{1}{2}%
\Pi\left[ 3H-W\frac{\dot{\rho}}{\rho}\right] \right) =0,
\end{equation}
and is given by
\begin{equation}
\left( \frac{(\omega+1)^{2}}{k_{\gamma}^{2}}\rho^{1-2\gamma}-3\left( 1-\frac{%
(\omega+1)}{2}\right) ^{2}\right) \rho=3\Lambda\left( 1-\frac{(\omega+1)}{2}%
\right) ^{2},
\end{equation}
or alternately
\begin{equation}
\rho\left( k_{\gamma}^{-2}(\omega+1)^{2}\rho^{1-2\gamma}-\beta\right)
=\beta\Lambda,
\end{equation}
where $\beta=3\left( 1-\frac{(\omega+1)}{2}\right) ^{2}.$ As we can see
\begin{equation}
\rho=\left( k_{\gamma}^{2}\frac{\beta}{(\omega+1)^{2}}\right) ^{\frac
{1}{1-2\gamma}}=0\Longleftrightarrow\gamma=\frac{1}{2},
\end{equation}
since $\frac{\beta}{(\omega+1)^{2}}<1,$ $\forall\omega.$ For this value, we
have $\rho=\Lambda=0.$

Therefore, as has already been pointed out earlier$^{24-26}$,
causal bulk viscous cosmological models with $\gamma=1/2$ describe a
two-phase evolution of the Universe. The Universe is born from a zero energy
density vacuum state and in the first phase the energy density is increasing
to a maximum value. This period corresponds to a phase of matter creation.
After reaching a maximum value, due to the expansion of the Universe, the
energy density becomes a monotonically decreasing function of time and the
second phase describes a standard evolution of the bulk viscous causal
cosmological fluid. Hence, in this framework, bulk viscous processes can
model matter creation in the early Universe. The ultra-stiff case of the
Zel'dovich fluid, $\omega=1$ also gives $\rho=0 $, but from this value
we cannot find the value for $\gamma.$

The structurally stable approximation of the system perturbed by the viscous
parameter must have the same critical points as the unperturbed system. The
unperturbed system has no critical points except the point $(H,\rho)=(0,0)$.
Therefore it is observed that the only structurally stable approximations to
the flat FRW solutions are those with $\xi=k_{\gamma}\rho^{1/2}$, i.e. $%
\gamma=1/2$.

As already have been pointed out by M. Szydlowski and M. Heller$^{51}$
in the study of non-causal viscous fluids it is possible to use a
Lie group technique in order to determine an adequate equation of state, in
particular theses authors find that there is a only one symmetry for the
field equations if $\gamma=1/2.$

\section{A naive method\label{DA}}

In this section, we use dimensional analysis to obtain a complete set of
solutions for the field equations with the assumptions $div(T)=0$ and
$\Pi=\varkappa\rho$ and $\varkappa\in\mathbb{R}^{-}$. This leads to
\begin{equation}
\rho=A_{\omega,\varkappa}f^{-3\left( \omega+1+\varkappa\right) }\mbox{ \ \ \
or \ \ \ \ \ }\rho=A_{\omega,\varkappa}f^{-\alpha},
\end{equation}
valid for $\forall\gamma.$ The set of governing parameters are $%
\mathfrak{M=M}\left\{ A_{\omega,\varkappa},k_{\gamma},t\right\} ,$ where $\left[
A_{\omega,\varkappa}\right] =L^{3\left( \omega+1+\varkappa\right)
-1}MT^{-2}, $ $\left[ k_{\gamma}\right] =L^{-1/2}M^{1/2}$ and $\left[ t%
\right] =T.$ Using these, we obtain the following relations:
\begin{eqnarray}
G & \propto & A_{\omega,\varkappa}^{\frac{2}{\alpha+1}}k_{\gamma}^{\frac{%
3+\alpha}{b\left( \alpha+1\right) }}t^{-4-\frac{3+\alpha}{b\left(
\alpha+1\right) }},  \label{g} \\
c & \propto & A_{\omega,\varkappa}^{\frac{1}{\alpha+1}}k_{\gamma}^{\frac
{1}{b\left( \alpha+1\right) }}t^{-1-\frac{1}{b\left( \alpha+1\right) }},
\label{c} \\
\rho & \propto & k_{\gamma}^{-b^{-1}}t^{b^{-1}}\propto\Pi,  \label{d} \\
\xi & = & k_{\gamma}\rho^{\gamma},\mbox{ \ \ }\tau=\xi\rho^{-1}  \label{spi} \\
f & \propto & A_{\omega,\varkappa}^{\frac{1}{\alpha+1}}k_{\gamma}^{\frac
{1}{b\left( \alpha+1\right) }}t^{-\frac{1}{b\left( \alpha+1\right) }},
\label{f} \\
k_{B}\theta & \propto & A_{\omega,\varkappa}^{\frac{3}{\alpha+1}}k_{\gamma }^{%
\frac{3}{b\left( \alpha+1\right) }-\frac{1}{b}}t^{-\frac{3}{b\left(
\alpha+1\right) }+\frac{1}{b}},  \label{temp} \\
a^{-1/4}s & \propto & A_{\omega,\varkappa}^{\frac{3}{\alpha+1}}k_{\gamma }^{%
\frac{3}{4}\left( \frac{3-\alpha}{b\left( \alpha+1\right) }\right) }t^{-%
\frac{3}{4}\left( \frac{3-\alpha}{b\left( \alpha+1\right) }\right) },
\label{entro} \\
\Lambda & \propto & A_{\omega,\varkappa}^{\frac{-2}{\alpha+1}}k_{\gamma}^{%
\frac{-2}{b\left( \alpha+1\right) }}t^{\frac{2}{b\left( \alpha+1\right) }},
\label{lambda} \\
q & = & -b\alpha-b-1  \label{q}
\end{eqnarray}
where, $\alpha=3(\omega+1+\varkappa)-1,$ $b=\gamma-1,$ $s$ is the entropy
and $q$ is the deceleration parameter. Note that the current method is very
different from the method employed in the previous section.

It is observed that
\begin{equation}
\frac{G}{c^{2}}=const.\Longleftrightarrow b=-\frac{1}{2}\Longleftrightarrow
\gamma=\frac{1}{2},
\end{equation}
and for $\gamma=\frac{1}{2}$, we obtain the following results:
\begin{eqnarray}
G & \propto & t^{-2\left( \frac{\alpha-1}{\alpha+1}\right) },\mbox{ \ \ }%
c \propto t^{-\left( \frac{\alpha-1}{\alpha+1}\right) },\mbox{ \ \ \ \ }%
\rho\propto t^{-2}\propto\Pi, \\
\mbox{ \ }f & \propto & t^{\frac{2}{\left( \alpha+1\right) }},\mbox{ \ \ \ }%
k_{B}\theta\propto t^{-2\left( \frac{\alpha-2}{\alpha+1}\right) },\mbox{ \ \
}a^{-1/4}s\propto t^{\frac{3}{2}\left( \frac{3-\alpha}{\left(
\alpha+1\right) }\right) }, \\
\Lambda & \propto & t^{\frac{-4}{\left( \alpha+1\right) }},\mbox{ \ \ \ }q=%
\frac{\alpha-1}{2}.
\end{eqnarray}
Thus, we see that (for $\gamma=\frac{1}{2})$ $G, c$ and $\Lambda$ are
decreasing functions on time despite the fact that the relation $G/c^{2}$
remains constant. In a previous work$^{52}$, where we only consider $G$
and $\Lambda$ as time functions, we found that for $\gamma=1/2$, $G$ behaves
as a true constant i.e. $G=const.$ and that $\Lambda$ vanishes. In this new
scenario, we find that $G$ and $c$ varies with time but in such a way that
the relation $G/c^{2}=const.$ and $\Lambda$ is a decreasing funtion on time.
This consequence has some interesting implications in light of recent
supernovae observations which suggest a small but non-zero value for the
cosmological constant\cite{Ca92}.

We would like to point out that these results are very similar to those in
$^{54}$ where a perfect fluid cosmological model with time-varying
constants in the presence of adiabatic matter creation was considered. This
striking similarity of our results is not surprising as it has been
suggested by Zeldovich$^{55}$ and later by Murphy$^{56}$ and Hu$^{57}$
that the introduction of viscosity in the cosmological fluid is a
phenomenological description of the effect of creation of particles by the
non-stationary gravitational field of the expanding cosmos. A non-vanishing
particle production rate is equivalent to a bulk viscous pressure in the
cosmological fluid or from a quantum point of view, with a viscosity of the
vacuum. This is due to the simple circumstance that any source term in the
energy balance of a relativistic fluid may be formally rewritten in terms of
an effective bulk viscosity. Zimdahl and others$^{40}$ have considered in detail
the possibility that the bulk viscous pressure of the full
Israel-Stewart-Hiscock theory may also be interpreted as an effective
description for particle production processes. The particle creation process
leads to considerable changes in the thermodynamical behavior of the
Universe. If the chemical potential of the newly created particles is zero, $%
\mu=0$, then the non-vanishing bulk pressure $\Pi$ associated with an
increase in the number of fluid particles satisfies formally the same
equation as in the case of presence of a real dissipative bulk viscosity$^{24-33}$.

This coincidence can be easily explained by observing the set of the
governing parameters. In the case of a viscous fluid, this set is
characterized by $\mathfrak{M=M}\left\{
A_{\omega,\varkappa},k_{\gamma},t\right\} ,$ where $\left[
A_{\omega,\varkappa}\right] =L^{3\left( \omega+1+\varkappa\right)
-1}MT^{-2}, $ $\left[ k_{\gamma}\right] =L^{-1/2}M^{1/2}$ and $\left[ t%
\right] =T,$ while in the case of adiabatic matter creation, this set is
characterized by $\mathfrak{M=M}\left\{ A_{\omega,\beta},B,t\right\} $ with $%
\left[ A_{\omega,\beta}\right] =L^{3(\omega+1)(1-\beta)-1}MT^{-2}$, $\left[
G/c^{2}=B\right] =LM^{-1}$ and $\left[ t\right] =T,$ see$^{41}$ 
for the details. We can observe the extreme similarity between $%
A_{\omega,\varkappa}$ and $A_{\omega,\beta}$ and the coincidence between $B$
and $k_{\gamma}^{-2}$, which has been shown in the previous section as $%
k_{\gamma}^{2}=c^{2}/G$.

Another similarity of the present results with those obtained in$^{54}$ 
is that for adiabatic matter creation, we cannot impose such a
mechanism with $\omega= 0$ on the equation of state for matter predominance.
In the same way, we observe from eq. (\ref{temp}) that if we fix $\omega=0$
into the equation of state the fluid temperature increases indefinitely with
time. Thus, we arrive at a similar conclusion that, in the matter
predominance era, we cannot consider the effects of the bulk viscosity in
our fluid. This fact is in agreement with a previous result obtained in$^{58}$ 
where we studied a full causal bulk viscous model through the
renormalization group method arriving to the conclusion that in the
long-time asymptotics the cosmological model tends to a flat ideal
(non-viscous) Friedmann type geometry.

We also notice that the $\varkappa-$parameter, i.e. the causal bulk viscous
effect, weakly perturbs the FRW perfect fluid solution in such a way that
the comoving entropy varies with $t,$ while in the perfect fluid case, i.e. $%
\varkappa=0$, the comoving entropy is constant. The same considerations
could be taken into account for the scale factor $f(t)=f_{0}t^{\frac{1}{%
3\left( 1-\gamma\right) \left( \omega+1+\varkappa\right) }},$ which is
perturbed weakly for the bulk viscous parameter as we can see in the special
case of $\gamma=1/2.$ The solutions obtained for the scale factor, energy
density, entropy, etc. are similar ``but different'' to those obtained for a
perfect fluid model where the $\varkappa-$parameter vanishes. We would like
to point out that our model is thermodynamically consistent for the usual
matter equations of state and valid for all $\gamma-$parameter, i.e. $%
\forall \gamma\in\left( 0,1\right) .$ In the same way, we can see that the
viscous parameter helps us to get rid of the so-called entropy problem since
in this model entropy varies with $t.$

The bulk viscous pressure and the energy density of the cosmological fluid
are proportional to each other and hence, the general evolution of $\Pi$ is
qualitatively similar to the evolution of the thermodynamic pressure $p$,
both obeying a similar equation of state. In fact, by defining an effective
coefficient $\omega_{eff}=\omega-\varkappa$, the equation of state of the
cosmological fluid can be formally written as $p_{eff}=p+\Pi=\left(
\omega_{eff}-1\right) \rho$. However, since the bulk viscous pressure of the
cosmological fluid must also satisfy the evolution equation (\ref{nfield5}),
the resulting time evolution depends not only on $\omega_{eff}$, but also,
via the coefficients $\delta$ and $\gamma$, on the equations of state of the
bulk viscosity coefficient, $\xi=\xi\left( \rho\right) $ and of the
temperature, $T=T\left( \rho\right) $. Therefore, even that formally one can
introduce an effective pressure (including both $p$ and $\Pi$), obeying a $%
\omega$-law equation of state, due to the extra-constraints imposed by the
requirements of the causal bulk evolution, the general dynamics of the
present model cannot be reduced to a perfect fluid model evolution.

\section{Direct integration of the field equations\label{INT}}

As we have seen in the previous section, dimensional analysis suggests us
that $\Pi\propto\rho.$ We formalize this relationship in the following way, $%
\Pi=\varkappa\rho$, and under physical considerations we take $\varkappa \in%
\mathbb{R}^{-}$ and $\left| \varkappa\right| \ll1$ $^{40}$. In this
section, we integrate the field equations with these assumptions and taking
into account the conservation principle. This leads us to the following
relationship
\begin{equation}
\dot{\rho}+3\left( \omega+1+\varkappa\right) \rho H=0.
\end{equation}

This trivially lead us to the well known relationship between the energy
density $\rho$ and the scale factor $f$
\begin{equation}
\rho=A_{\omega}f^{-3\left( \omega+1+\varkappa\right) }\mbox{ \ \ \ or \ \ \
\ \ }\rho=A_{\omega}f^{-\alpha},  \label{INT conservation}
\end{equation}
where $\alpha=3\left( \omega+1+\varkappa\right)$. Now, taking into account
the eq. (\ref{field4}) \ and simplifying it, we obtain,
\begin{equation}
\varkappa\dot{\rho}+\frac{\varkappa\rho}{k_{\gamma}\rho^{\gamma-1}}=\frac
{3}{\alpha}\dot{\rho}-\frac{1}{2}\varkappa\dot{\rho}\left( -\frac{3}{\alpha }%
-W\right) ,
\end{equation}
thereby obtaining $\rho=\rho(t).$ On simplifying further, we obtain,
\begin{equation}
\frac{\dot{\rho}}{\rho^{2-\gamma}}=\frac{K}{k_{\gamma}}\mbox{ \ \ \ }%
\Longrightarrow\mbox{ \ \ \ }\rho=dk_{\gamma}^{-b}t^{b},  \label{INT density}
\end{equation}
where
\begin{equation}
\mbox{\ \ \ }K=\left( \frac{\varkappa}{\frac{3}{\alpha}+\frac{3\varkappa }{%
2\alpha}+\frac{W\varkappa}{2}-\varkappa}\right) ,\mbox{ \ }d=\left(
\gamma-1\right) K^{b}\mbox{\ and \ \ \ \ }b=\frac{1}{\gamma-1}.
\end{equation}

From equation\ (\ref{INT conservation}) we obtain:
\begin{equation}
f=\left( \frac{A_{\omega}}{d}k_{\gamma}^{b}t^{-b}\right) ^{1/\alpha},\mbox{
\ \ i.e. \ \ }f\propto t^{\frac{-1}{3\left( \omega+1+\varkappa\right) \left(
\gamma-1\right) }}.  \label{INT sfactor}
\end{equation}
An important observational quantity is the deceleration parameter $q=\frac
{d}{dt}\left( \frac{1}{H}\right) -1$. The sign of the deceleration parameter
indicates whether the model inflates or not. The positive sign of $q$
corresponds to ``standard'' decelerating models whereas the negative sign
indicates inflation. In our model, the deceleration parameter behaves as:
\begin{equation}
q=-1-\frac{\alpha}{b}.
\end{equation}
Thus, for the parameters $\gamma, \omega$ and $\varkappa$, the deceleration
parameter indicates an inflationary behaviour when $\alpha/b > 0$ and a
non-inflationary behaviour for $\alpha/b < -1$. We proceed with the
calculation of the other physical quantities as under:
\begin{eqnarray}
\xi & = & k_{\gamma}\rho^{\gamma}\propto k_{\gamma}\left(
dk_{\gamma}^{-b}t^{b}\right)
^{\gamma}=d^{\gamma}k_{\gamma}^{-b\gamma}t^{\gamma b}, \\
T & = & D_{\delta}\rho^{\delta}=D_{\delta}\left( dk_{\gamma}^{-b}t^{b}\right)
^{\delta}\mbox{ \ \ \ with \ \ }\delta=\frac{\omega}{\omega+1}, \\
\tau & = & \xi\rho^{-1}=k_{\gamma}\left( dk_{\gamma}^{-b}t^{b}\right) ^{\left(
\gamma-1\right) },\mbox{ i.e. }\tau=d^{\gamma-1}t
\end{eqnarray}
We see from $\tau=d^{\gamma-1}t$ that this result is in agreement with the
theoretical result obtained in$^{35}$. For viscous expansion to be
non-thermalizing, we should have $\tau<t,$ or otherwise the basic
interaction rate for viscous effects should be sufficiently rapid to restore
the equilibrium as the fluid expands. The comoving entropy is
\begin{equation}
\Sigma(t)-\Sigma\left( t_{0}\right) =-\frac{3\varkappa\beta}{d^{\frac
{3b}{\alpha}+\delta}}\frac{A_{\omega}^{1/\alpha}}{k_{B}D_{\delta}}k_{\gamma
}^{b(-1+\frac{3}{\alpha}+\delta)}\left[ \tilde{t}^{b(1-\frac{3}{\alpha }%
-\delta)}\right] _{t_{0}}^{t}.
\end{equation}
We notice that the parameter $\varkappa$ weakly perturbs the perfect fluid
FRW Universe. When $\varkappa = 0$, the comoving entropy assumes a constant
value and we recover the perfect fluid case. Finally, we will use the
equations
\begin{eqnarray}
3H^{2} & = & \frac{8\pi G}{c^{2}}\rho+\Lambda c^{2},  \label{sil1} \\
\frac{\dot{\Lambda}c^{4}}{8\pi G\rho}+\frac{\dot{G}}{G}-4\frac{\dot{c}}{c}
& = & 0,  \label{sil2}
\end{eqnarray}
to obtain the behaviour of the ``constants'' $G,c$ and $\Lambda.$ From (\ref
{sil1}), we obtain $\dot{\Lambda}$ and using it in eq. (\ref{sil2}) we
obtain
\begin{equation}
-\frac{3\dot{c}c}{4\pi G\rho}H^{2}+\frac{c^{2}3H{\dot{H}}}{4\pi
G\rho}-\frac{\dot{\rho}}{\rho}=0  \label{flor}
\end{equation}
Considering the previous results
\begin{equation}
H=\beta t^{-1\mbox{ }}\mbox{ \ \ and \ \ \ }\rho=dk_{\gamma}^{-b}t^{b},
\end{equation}
where $\beta=-b/\alpha.$

These results are used in eq. (\ref{flor}) to obtain the next ODE
\begin{equation}
\frac{3\beta^{2}}{4\pi d}\frac{c^{2}t^{-2-b}}{Gk_{\gamma}^{-b}}\left[ \frac{%
\dot{c}}{c}+\frac{1}{t}\right] + b\frac{1}{t}=0.  \label{cova}
\end{equation}
We can see that the above equation verifies the relationship (a particular
solution)
\begin{equation}
\frac{k_{\gamma}^{b}c^{2}}{G}=gt^{b+2}\mbox{ \ \ \ \ \ therefore \ }%
G=gk_{\gamma}^{b}c^{2}t^{-b-2},
\end{equation}
$g$ being a numerical constant. It is observed that if $b=-2$ i.e. $%
\gamma=1/2$, we obtain $G/c^{2}=B$. We use this relation in eq.(\ref{cova})
obtaining
\begin{equation}
\frac{\dot{c}}{c}=-\left( 1+\frac{b}{K_{c}}\right) t^{-1},
\end{equation}
where $K_{c}=\frac{3\beta^{2} g}{4\pi d}$, and therefore,
\begin{equation}
c=\tilde{K}_{c}t^{\kappa},
\end{equation}
where $\tilde{K}_{c}$ is an integration constant and $\kappa=\left( -1-\frac{
b}{K_{c}}\right) .$

If we impose a new assumption (which is supported by observational
considerations of $\Lambda$ decaying with time and assuming a small but
non-zero value) as
\begin{equation}
\Lambda=\frac{l}{c^{2}(t)t^{2}}\mbox{ }\Longrightarrow\mbox{ \ }\dot{\Lambda
}=-\frac{2l}{c^{2}t^{2}}\left( \frac{\dot{c}}{c}+\frac{1}{t}\right) ,
\end{equation}
we can see from eq. (\ref{sil1}) that
\begin{equation}
\frac{G}{c^{2}}=\left[ \frac{3\beta^{2}-l}{8\pi d}\right] k_{%
\gamma}^{b}t^{-b-2}=Bk_{\gamma}^{b}t^{-b-2}.
\end{equation}
Taking all these relations into consideration, we use in eq.(\ref{sil2}) to
obtain
\begin{equation}
-\frac{l}{4\pi B}\left( \frac{\dot{c}}{c}+\frac{1}{t}\right) -2\frac{\dot {c}%
}{c}-\frac{b+2}{t}=0,
\end{equation}
whose trivial solution is
\begin{equation}
c=\mathfrak{K}_{c}t^{-\mu},
\end{equation}
where $\mathfrak{K}_{c}$ is an integration constant and $\mu=\left( \frac{%
l+4\pi K(b+2)}{l+8\pi K}\right) .$

We have solved our model under the assumptions: $div(T)=0,$ $\Pi=\varkappa
\rho$, with $\varkappa\in\mathbb{R}^{-}$ and valid for $\forall\gamma.$ It
is observed that if $\gamma=1/2$ we obtain $G/c^{2}=const.$ as obtained
earlier. Thus with the imposed hypotheses, we have obtained similar results.

\section{Lie method\label{LIE}}

In this section, we study the field equations using the symmetry method. As
we have discussed earlier, dimensional analysis is just a manifestation of
scaling symmetry. However, this type of symmetry is not the most general
form of symmetries$^{59-61}$.
Therefore, by studying the form of $G(t)$ and $c(t)$ for which the equations admit
symmetries, we expect to uncover new integrable models. With these
assumptions, we shall see that the solutions obtained in previous sections
are recovered.

We start with the assumption $\Pi=\varkappa\rho$, with $\varkappa\in \mathbb{%
R}^{-}$ The bulk viscosity evolution equation can then be rewritten in the
alternative form
\begin{equation}
\delta\frac{\dot{\rho}}{\rho}+k_{\gamma}^{-1}\rho^{1-\gamma}=3\beta H,
\end{equation}
where $\beta=\left( \frac{1}{\varkappa}-\frac{1}{2}\right) $ and $%
\delta=\left( 1-\frac{W}{2}\right) .$

Taking the derivative with respect to the time of this equation and with the
use of the next equation

\begin{equation}
\dot{H}=-4\pi\alpha\frac{G(t)}{c^{2}(t)}\rho,
\end{equation}
obtained from the field equations (where $\alpha=\left( 1+\omega
+\varkappa\right) $), we obtain the following second order differential
equation describing the time variation of the density of the cosmological
fluid:
\begin{equation}
\ddot{\rho}=\frac{\dot{\rho}^{2}}{\rho}-A\rho^{s}\dot{\rho}+B\frac{G(t)}{%
c^{2}(t)}\rho^{2},  \label{neweq}
\end{equation}
where $A=\frac{k_{\gamma}^{-1}(1-\gamma)}{\delta},$ $s=(1-\gamma)$ and $B=%
\frac{12\pi\alpha\beta}{\delta}.$

Equation (\ref{neweq}) is of the general form.
\begin{equation}
\ddot{\rho}=\psi(t,\rho,\dot{\rho}),
\end{equation}
where $\psi(t,\rho,\dot{\rho})=\frac{\dot{\rho}^{2}}{\rho}-A\rho^{s}\dot{%
\rho }+B\frac{G(t)}{c^{2}(t)}\rho^{2}$.

We are going now to apply all the standard procedure of Lie group analysis
to this equation (see$^{59}$ for details and notation)

A vector field $X$
\begin{equation}
X=\xi(t,\rho)\partial_{t}+\eta(t,\rho)\partial_{\rho},
\end{equation}
is a symmetry of (\ref{neweq}) if
\begin{eqnarray}
-\xi\psi_{t}-\eta\psi_{\rho}+\eta_{tt}+\left( 2\eta_{t\rho}-\xi_{tt}\right)
\dot{\rho}+\left( \eta_{\rho\rho}-2\xi_{t\rho}\right) \dot{\rho}%
^{2}-\xi_{\rho\rho}\dot{\rho}^{3}+ \nonumber
\end{eqnarray}
\begin{equation}
+\left( \eta_{\rho}-2\xi_{t}-3\dot{\rho}\xi_{\rho}\right) \psi-\left[
\eta_{t}+\left( \eta_{\rho}-\xi_{t}\right) \dot{\rho}-\dot{\rho}%
^{2}\xi_{\rho}\right] \psi_{\dot{\rho}}=0.  \label{ber2}
\end{equation}

By expanding and separating (\ref{ber2}) with respect to powers of $\dot{%
\rho }$ we obtain the overdetermined system:
\begin{eqnarray}
\xi_{\rho\rho}+\rho^{-1}\xi_{\rho} & = & 0,  \label{ber1} \\
\eta\rho^{-2}-\eta_{\rho}\rho^{-1}+\eta_{\rho\rho}-2\xi_{t\rho}+2A\xi_{\rho
}\rho^{1-s} & = & 0,  \label{ber1_1} \\
2\eta_{t\rho}-\xi_{tt}+A\xi_{t}\rho^{s}-3B\xi_{\rho}\frac{G}{c^{2}}\rho
^{2}-2\eta_{t}\rho^{-1}+As\eta\rho^{s-1} & = & 0,  \label{ber3} \\
-B\xi\left( \frac{\dot{G}}{c^{2}}-2G\frac{\dot{c}}{c^{3}}\right) \rho
^{2}-2B\eta\frac{G}{c^{2}}\rho+\eta_{tt}+\left( \eta_{\rho}-2\xi_{t}\right) B%
\frac{G}{c^{2}}\rho^{2}+A\eta_{t}\rho^{1-s} & = & 0,  \label{ber4}
\end{eqnarray}
Solving (\ref{ber1}-\ref{ber4}), we find that
\begin{equation}
\xi(\rho,t)=-ast+b,\mbox{\ \ \ \ \ \ \ \ }\eta(\rho,t)=a\rho,
\end{equation}
with the constraint
\begin{equation}
\frac{\dot{G}}{G}=2\frac{\dot{c}}{c}+\frac{(1-2s)a}{b-ast},
\end{equation}
$a$ and $b$ being numerical constants.

Thus, we have found all the possible forms of $G$ and $c$ for which eq. (\ref
{neweq}) admits symmetries. There are two cases with respect to the values
of the constant $a$; $a=0$ which correspond to $G=const,$ $c=const.$ and $%
a\neq0$ which correspond to
\begin{equation}
\frac{G}{c^{2}}=K\left[ ast+b\right] ^{-2+\frac{1}{s}},  \label{SYM G}
\end{equation}
with $K$ a constant of integration. If $s=\frac{1}{2},$ which corresponds to
$\gamma=\frac{1}{2},$ then $G$ and $c$ remain constants. For this form of $G$
and $c$ eq. (\ref{neweq}) admits a single symmetry
\begin{equation}
X=\left( ast-b\right) \partial_{t}-\left( a\rho\right) \partial_{\rho}.
\label{SYM X}
\end{equation}

The knowledge of one symmetry $X$ might suggest the form of a particular
solution as an invariant of the operator $X$, i.e. the solution of
\begin{equation}
\frac{dt}{\xi\left( t,\rho\right) }=\frac{d\rho}{\eta\left( t,\rho\right) }.
\label{ecu7}
\end{equation}

This particular solution is known as an invariant solution (generalization
of similarity solution). In this case
\begin{equation}
\rho=\rho_{0}t^{-\frac{1}{s}},  \label{SYM density}
\end{equation}
where for simplicity we have taken $b=0,$ and $\rho_{0}$ is a constant of
integration (note that $s=\left( 1-\gamma\right) $ being $\gamma$ the bulk
viscous parameter).

We can apply a pedestrian method to try to obtain the same results. In this
way, taking into account dimensional considerations, from the eq. (\ref
{neweq}) we obtain the following relationships between density, time and
gravitational constant:
\begin{equation}
k_{\gamma}^{-1}s\rho^{s}t\backsimeq1,\mbox{ \ \ }B\frac{G}{c^{2}}\rho
t^{2}\backsimeq1.  \label{sciama}
\end{equation}

This last relationship is also known as the relation for inertia obtained by
Sciama. From these relationships, we obtain
\begin{equation}
\rho\thickapprox B^{-1}K^{-1}c^{2}\left[ ast+b\right] ^{\frac{2s-1}{s}%
}t^{-2}\thickapprox B^{-1}K^{-1}c^{2}\left[ ast\right] ^{-\frac{1}{s}}.
\label{sciama density}
\end{equation}
We can see that this result verifies the relation
\begin{equation}
k_{\gamma}^{-1}s\rho^{s}t\backsimeq1.\mbox{ \ \ }
\end{equation}
Once we have obtained $\rho$, we can obtain $f$ (the scale factor) from
\begin{equation}
\rho=A_{\omega,\varkappa}f^{-3\left( \omega+1+\varkappa\right)
}\Longrightarrow f=\left( A_{\omega,\varkappa}\rho_{0}^{-1}t\right) ^{\frac{1%
}{3\left( \omega+1+\varkappa\right) s}},
\end{equation}
In this way, we find $H$ and from eq. (\ref{nfield2}) the behaviour of $%
\Lambda$ is obtained as:
\begin{equation}
\Lambda=\frac{\left( 3\kappa^{2}-8\pi K(as)^{-2+\frac{1}{s}}\rho_{0}\right)
}{c^{2}t^{2}}=\frac{l}{c^{2}t^{2}},  \label{lambda2}
\end{equation}
where, $\kappa=\frac{\left( A_{\omega,\varkappa}\rho_{0}^{-1}\right) }{%
3\left( \omega+1+\varkappa\right) s}$. On replacing all these results into
eq. (\ref{nfield4}), we obtain the exact behaviour for $c,$ i.e.
\begin{equation}
-\left( \frac{l}{4\pi\widetilde{K}\rho_{0}}+\frac{1-2s}{s}\right) \frac
{1}{t}=\left( \frac{l}{4\pi\widetilde{K}\rho_{0}}+2\right) \frac{\dot{c}}{c},
\end{equation}
therefore
\begin{equation}
c=c_{0}t^{-\varepsilon},
\end{equation}
where, $\varepsilon=\frac{\left( \lambda+\frac{(1-2s)}{s}\right) }{\left(
\lambda+2\right) }$ and $\lambda=\frac{l}{4\pi\widetilde{K}\rho_{0}}.$

Therefore, the assumption $\Pi=\varkappa\rho$ together with the equation (%
\ref{nfield5}) is very restrictive. Hence, we conclude that under these
assumptions, the field equations do not admit any other solution.
\section{Relaxing the hypotheses\label{REL}}

In the previous section, we have seen that our assumption $\Pi=\varkappa\rho$
together with the equation (\ref{nfield5}) does not give rise to any new
solutions. This motivates us to investigate our assumptions further and
relax the hypotheses we have used in obtaining solutions to the field
equations. With the following assumptions obtained (suggested) from the
sub-section \ref{dym}, we would like to obtain an equation that helps us to
determine the behaviour of the ``constants''. The new hypotheses are:

\begin{enumerate}

\item

Relationship between $G$ and $c$:
\begin{equation}
\frac{G}{c^{2}}=Kk_{\gamma}^{b}H^{2+b}.
\end{equation}

\item
$\Lambda$ follows the law:
\begin{equation}
\Lambda=\frac{lH^{2}}{c^{2}}  \label{Lambda1}
\end{equation}

\end{enumerate}

where $K,$ $l\in\mathbb{R}$ are numerical constants and $b=1/(\gamma-1)$.

In this way, we allow the scale factor $f$ to be an arbitrary function of
time and it does not follow a power law i.e. $f\propto t^{\alpha} $ as in
previous sections. From the eq. (\ref{sil1}) and taking into account the
assumptions, we obtain
\begin{equation}
\rho=dH^{-b},
\end{equation}
where, $d=\left[ \frac{3-l}{8\pi Kk_{\gamma}^{b}}\right] .$ Therefore from (%
\ref{nfield3}), we obtain
\begin{equation}
\Pi=\frac{bd}{3}H^{-2-b}\dot{H}-d\left( \omega+1\right) dH^{-b}
\label{jana}
\end{equation}
If we introduce these results into the equation (\ref{nfield5}), then
\begin{equation}
{\ddot{H}}+K_{1}H{\dot H}+K_{2}H^{-1}\dot{H}^{2}+K_{3}H^{3}=0  \label{analy1}
\end{equation}
where $W=\left( \frac{2\omega+1}{\omega+1}\right) $ and
\begin{equation}
K_{1}=\left( 3+k_{\gamma}^{-1}d^{\gamma-1}\right) , \mbox{ \ }K_{2}=\left(
-b-2+Wb\right) \mbox{ \ }, K_{3}=3\left( \frac{9+3\omega}{2b}-\frac
{k_{\gamma}^{-1}d^{\gamma-1}\left( \omega+1\right) }{b}\right) ,
\end{equation}
In order to solve eq. (\ref{analy1}), we use the Lie method which results in
the following overdetermined system:
\begin{eqnarray}
-\xi_{HH}+\xi_{H}K_{2}H^{-1} & =0  \label{def1} \\
-\eta K_{2}H^{-2}+\eta_{HH}-2\xi_{tH}+\eta_{H}K_{2}H^{-1}+2\xi_{H}K_{1}H & =0
\label{def2} \\
\eta
K_{1}+2\eta_{tH}-\xi_{tt}+\xi_{t}K_{1}H+3\xi_{H}K_{3}H^{3}+2%
\eta_{t}K_{2}H^{-1} & =0  \label{def3} \\
3\eta
K_{3}H^{2}+\eta_{tt}-\eta_{H}K_{3}H^{3}+2\xi_{t}K_{3}H^{3}+\eta_{t}K_{1}H &
=0  \label{def4}
\end{eqnarray}
Solving (\ref{def1}-\ref{def4}), we find that
\begin{equation}
\xi(t,H)=-at+b,\mbox{\ }\eta(t,H)=aH
\end{equation}
$a$ and $b$ being numerical constants. For this form of $\xi$ and $\eta$,
eq. (\ref{analy1}) admits a single symmetry
\begin{equation}
X=(at-b)\partial_{t}-\left( aH\right) \partial_{H}.
\end{equation}

Following the standard procedure, we find
\begin{equation}
\frac{dt}{\xi\left( t,H\right) }=\frac{dH}{\eta\left( t,H\right) }%
\Longrightarrow H=\frac{1}{at-b},
\end{equation}
which imply that
\begin{equation}
f=\left( at-b\right) ^{\frac{1}{a}}  \label{scaleH}
\end{equation}
Thus, we obtain a power law solution as in previous section. To complete the
set of equations, we have to take into account the next equation
\begin{equation}
-\frac{\dot{\Lambda}c^{4}}{8\pi G\rho}-\frac{\dot{G}}{G}+4\frac{%
\dot{c}}{c}=0  \label{variation}
\end{equation}
and replace it into all our results thereby leading to the next
relationship,
\begin{equation}
c=KH^{\alpha}  \label{ch}
\end{equation}
where $\alpha=\frac{l+A(2+b)}{l+2A}$ and $\ A=4\pi Kd.$ In this way, we can
see that
\begin{equation}
G=KH^{2\left( \alpha+1\right) +b}\mbox{ \ \ and \ \ \ \ }\Lambda
=lH^{2\left( 1-\alpha\right) }.  \label{gh}
\end{equation}

On using a straightforward calculation, we see that we have obtained the
same solution even with different assumptions. The hypotheses considered
earlier were relaxed in the above analysis with a view of obtaining a new
set of solutions. This suggests that solutions obtained earlier are of a
fairly general nature and do not really depend on any specific assumptions.

\section{Conclusions.\label{CON}}

In the present paper, we have studied a causal bulk viscous cosmological
model with time varying constants. The dimensional analysis of the model
allows us to make the following assumptions: the bulk viscous pressure is
proportional to the energy density of the cosmological fluid and that for $%
\gamma=1/2$, the relation $G/c^{2}$ remains constant in spite of considering
``constants'' $G$ and $c$ as functions of $t$. In addition, we have shown
that for $\gamma=1/2$, the dynamics of our model is similar to that a
perfect fluid FRW model.

With the help of these assumptions, the general solution of the
gravitational equations can be obtained in an exact form, leading to a
power-law behavour of the physical parameters on the cosmological time.
These solutions show us that the ``constants'' $G, c$ and $\Lambda$ are
decreasing functions of time and that for $\gamma=1/2$, we obtain a model
similar to a perfect fluid cosmological model with time-varying constants in
the presence of adiabatic matter creation. This striking similarity of our
results is not surprising as it has been pointed out by many authors that
the introduction of viscosity in the cosmological fluid is a
phenomenological description of the effect of creation of particles by the
non-stationary gravitational field of the expanding cosmos. We have shown
that we cannot consider the bulk viscosity in the matter predominance era
since the fluid temperature increases indefinitely with time. We arrive at
the same conclusion in the case of adiabatic matter creation. This fact is
in agreement with a previous result where we studied a full causal bulk
viscous model through the renormalization group method arriving to the
conclusion that in the long-time asymptotics the cosmological model tends to
a flat ideal (non-viscous) Friedmann type geometry.

We also notice that the $\varkappa-$parameter, i.e. the causal bulk viscous
effect, weakly perturbs the FRW perfect fluid solution in such a way that
the comoving entropy varies with $t,$ while in the perfect fluid case, i.e. $%
\varkappa=0$, the comoving entropy is constant. The same considerations
could be taken into account for the scale factor $f(t)=f_{0}t^{\frac{1}{%
3\left( 1-\gamma\right) \left( \omega+1+\varkappa\right) }},$ which is
perturbed weakly for the bulk viscous parameter as we can see in the special
case of $\gamma=1/2.$ The solutions obtained for the scale factor, energy
density, entropy, etc. are similar ``but different'' to those obtained for a
perfect fluid model where the $\varkappa-$parameter vanishes. We would like
to point out that our model is thermodynamically consistent for the usual
matter equations of state and valid for all $\gamma-$parameter, i.e. $%
\forall \gamma\in\left( 0,1\right) .$ In the same way, we can see that the
viscous parameter helps us to get rid of the so-called entropy problem since
in this model entropy varies with $t.$

In the model considered, the evolution of the Universe starts from a
singular state, with the energy density, bulk viscosity coefficient and
cosmological constant tending to infinity. At the initial moment, $t=0$, the
relaxation time is $\tau=0$. From the singular initial state, the Universe
starts to expand, with the scale factor $f$ as a $\gamma$-dependent
function. The ``constants'' $G, c$ and $\Lambda$ are all decreasing
functions of time. In the radiation predominance era we can take into
account the effects of the bulk viscosity in all the quantities but when the
universe enters a matter predominance era, these effects vanish and in this
case our model behaves as a perfect fluid model with variable constants. The
present model is not defined for $s=1$, showing that in the important limit
of small densities a different approach is necessary.

The unicity of the solution is also proved by the investigation of the Lie
group symmetries of the basic equation describing the time variation of the
mass density of the Universe.

Bulk viscosity is expected to play an important role in the early evolution
of the Universe, when also the dynamics of the gravitational and
cosmological constants and the speed of light could be different. Hence the
present model, despite its simplicity, can lead to a better understanding of
the dynamics of our Universe in its first moments of existence. {\small {\ }}

\end{document}